\documentclass[twocolumn]{tMOP2e}

\usepackage{upgreek, graphicx, subfigure, amsmath}

\newcommand{\eref}[1]{Eq.~(\ref{#1})}
\newcommand{\erefs}[1]{Eqs.~(\ref{#1})}
\newcommand{\fref}[1]{Fig.~\ref{#1}}
\newcommand{\Sref}[1]{Section~\ref{#1}}

\newcommand{\ie}{i.e.}
\newcommand{\rmd}{~\text{d}}
\newcommand{\force}{\boldsymbol{F}}
\newcommand{\diffn}{\boldsymbol{D}}

\newcommand{\re}[1]{\,\text{Re}\!\left\{#1\right\}}

\makeatletter
\newlength \figurewidth
\makeatother

\begin{document}

\title{Amplified optomechanics in a unidirectional ring cavity}

\author{Andr\'e Xuereb,\textsuperscript{a,}$^\ast$\thanks{$^\ast$Corresponding author. Electronic address:~andre.xuereb@soton.ac.uk}
Peter Horak,\textsuperscript{b} and
Tim Freegarde\textsuperscript{a}\\
\vspace{6pt}
\textsuperscript{a}{School of Physics and Astronomy, University of Southampton, Southampton SO17~1BJ, United Kingdom};
\textsuperscript{b}{Optoelectronics Research Centre, University of Southampton, Southampton SO17~1BJ, United Kingdom}\\
\vspace{6pt}
\received{\today}}

\maketitle

\begin{abstract}
{We investigate optomechanical forces on a nearly lossless
scatterer, such as an atom pumped far off-resonance or a
micromirror, inside an optical ring cavity. Our model introduces
two additional features to the cavity: an isolator is used to
prevent circulation and resonant enhancement of the pump laser
field and thus to avoid saturation of or damage to the scatterer,
and an optical amplifier is used to enhance the effective
$Q$-factor of the counterpropagating mode and thus to increase the
velocity-dependent forces by amplifying the back-scattered light.
We calculate friction forces, momentum diffusion, and steady-state
temperatures to demonstrate the advantages of the proposed setup.}
\end{abstract}


\section{Introduction and Motivation}\label{sec:Introduction}
Free-space laser cooling~\cite{Chu1985} has proven to be
remarkably successful in cooling simple atomic systems,
{especially alkali} atoms~\cite{Chu1998}. {Relying} on the
availability of a (quasi-)closed two-level~\cite{Metcalf2003} or
multi-level~\cite{Dalibard1989,Ungar1989} system, however, only
occasional successes were had with more complicated systems, such
as molecules~\cite{Zeppenfeld2009,Shuman2010}. An alternative to
such schemes is cavity--mediated
cooling~\cite{Hechenblaikner1998,Maunz2004,Leibrandt2009,Koch2010},
where the interaction of a polarisable particle with a cavity
field leads to a Sisyphus--type mechanism~\cite{Horak1997} that
can cool the motion of the particle; no specific energy {level}
scheme is required for this mechanism to operate. {Much of this
work has been focused on standing-wave (Fabry--P\'erot) cavities,
where high $Q$-factors can be achieved experimentally to
significantly enhance optomechanical forces. However, in the limit
of strong scatterers, friction forces in standing-wave cavities
become increasingly position-dependent~\cite{Xuereb2010c}, which limits
the overall, averaged cooling efficiency. This can be overcome by
using ring
cavities~\cite{Gangl2000a,Elsasser2003,Kruse2003,Nagy2006,Slama2007,Hemmerling2010,Schulze2010,Niedenzu2010}
where the translational symmetry guarantees position-independent
forces. On the other hand, ring cavities are usually much larger
and of lower $Q$-factor than their standing-wave counterparts.
Using a gain medium inside a ring cavity has} been
proposed~\cite{Vuletic2001a,Salzburger2006} {to offset these
losses}, allowing one {to effectively `convert'} a low-$Q$
cavity into a high-$Q$ one, and {thus to} increase the
effective optomechanical interaction by orders of magnitude. This
{same concept} has also been raised in connection with using
optical parametric amplifiers in standard optomechanical
systems~\cite{Huang2009}. {Research is also being} conducted
into using nonlinear media inside cavities~\cite{Kumar2010} as a
tool to control the dynamics of a micromechanical oscillator.
{Another application of ring cavities is in the investigation
of collective atomic recoil lasing (CARL)~\cite{Bonifacio1994},
which exploits the spontaneous self-organisation of an atomic
ensemble within a ring cavity, induced by a strong pump beam, to
amplify a probe beam through Doppler-shifted reflection of the
pump. The gain medium is in this case the atomic ensemble itself.}
\par
{Here, we investigate a different system that shares several
features with the above mechanisms. In particular, we consider a
particle inside a ring cavity that includes a gain medium,
spatially separated from the particle. An isolator is also
included in the ring cavity, in such a way as to prevent the pump
beam from circulating in the cavity and being amplified;
this ensures that the intensity of the field surrounding the
particle is always low and thereby circumvents any problems caused
by atomic saturation or mirror burning. The Doppler-shifted
reflection of the pump from the particle is, on the other hand,
allowed to circulate, and its amplification in turn enhances the
velocity-dependent forces acting on the particle.}

 {In such a
situation}, one is able to take advantage of properties inherent
to the ring cavity system, such as the fact that the forces acting
on the particle do not exhibit any sub-wavelength spatial
modulation; this is due to the translational symmetry present in
the system~\cite{Gangl2000a}. Moreover, modest amplification
allows one to use optical fibres to form the ring cavity, opening
the door towards increasing the optical length of such cavities.
The optomechanical force is, {as we will see and} in the
parameter domain of interest, linearly dependent on the cavity
length; lengthening the cavity thus provides further enhancement
of the interaction.
\par
This paper is structured as follows. We shall first introduce the physical model, which we proceed to solve using the transfer matrix method (TMM)~\cite{Xuereb2009b} to obtain the friction force and momentum diffusion acting on the particle. In the good-cavity limit, \Sref{sec:GoodCavity}, simple expressions for these quantities can be obtained, yielding further insight into the system and allowing us to draw some parallels with traditional cavity cooling. {In this limit, our model becomes equivalent to one based on a standard master equation approach}~\cite{Gangl2000a} {as outlined} in \Sref{sec:Semiclassical}. Realistic numerical values for the various parameters are then used in \Sref{sec:Results} to explore the efficiency and limits of the cooling mechanism. Finally, we will conclude by mentioning some possible extensions to the scheme.

\begin{figure*}[t]
 \centering
 \includegraphics[width=\linewidth]{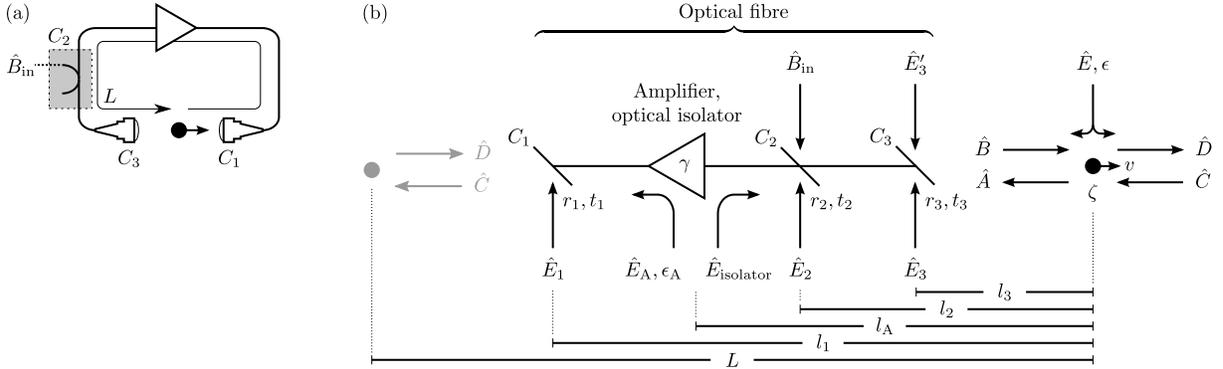}
  \caption{(a)~Physical schematic of a polarisable particle in a unidirectional ring cavity, showing the input field $\hat{B}_\text{in}$. (b)~Equivalent `transfer matrix'--style (unfolded) model; the particle is drawn on both sides of this schematic to illustrate the recursive nature of the cavity. The various components are defined in \Sref{sec:Model}.}
 \label{fig:Model}
\end{figure*}

\section{General expressions and equilibrium behaviour}\label{sec:Model}

The mathematical model of the ring cavity system, schematically
drawn in \fref{fig:Model}(a), is shown in \fref{fig:Model}(b). A
particle, characterised by its polarisability $\zeta$, is in a
ring cavity of round-trip length $L$. $C_{1,3}$ are the couplers,
between which the particle lies, that terminate the fibre-based
cavity, and $C_2$ is the input coupler that injects a pump beam
with wavenumber $k_0$ into one of the travelling modes of the
cavity. The couplers $C_i$ ($i=1,2,3$) have (amplitude) reflection
and transmission coefficients $r_i$ and $t_i=r_i+1$, respectively,
and are associated with corresponding noise modes $\hat{E}_i$.
Similarly, the particle itself couples a noise mode $\hat{E}$ into
the system with an amplitude $\epsilon$ depending on $\zeta$. The
length scales $l_1>l_\text{A}>l_2>l_3>0$ are introduced for
clarity, but their values are not important for the results of
this paper. The cavity contains an optical isolator which prevents
the pumped mode from circulating inside the cavity. This avoids
resonant enhancement of the pumped mode in the cavity and thus
avoids saturating the particle. The backscattered
counterpropagating mode, on the other hand, is amplified on every
round trip by means of an optical amplifier with gain $\gamma\geq
1$. The amplifier also introduces a noise mode $\hat{E}_\text{A}$
with an amplitude $\epsilon_\text{A}$ which depends on $\gamma$.
The TMM{~\cite{Xuereb2009b}} is used to self-consistently solve
for the two counterpropagating field amplitudes {at every point
in the cavity in the presence of the pump field and the noise
modes. Note that in the limit where the amplifier is compensating
for the ring cavity losses, the amplifier noise is also comparable
to the loss-induced noise and must therefore be} taken into
account in our model.
\par
Using the notation in \fref{fig:Model}(a) we can relate the expectation values of the amplitudes of the {two input and two scattered} field modes interacting with the particle {in a one-dimensional scheme}, {$A(k)=\langle\hat{A}(k)\rangle$, $B(k)=\langle\hat{B}(k)\rangle$, $C(k)=\langle\hat{C}(k)\rangle$, and $D(k)=\langle\hat{D}(k)\rangle$, to $B_\text{in}(k)=\langle\hat{B}_\text{in}(k)\rangle$} by means of the relations
\begin{equation}
\label{eq:BasicRelations}
B = r_2t_3e^{ik_0l_2}B_\text{in}\,,\ C=\alpha A\,,\ \text{and}\ \begin{pmatrix}A\\B\end{pmatrix}=\hat{M}\begin{pmatrix}C\\D\end{pmatrix}\,,
\end{equation}
where {$\alpha(k)=t_1t_2t_3\gamma\exp(ikL)$} is the factor multiplied to the field amplitude every round trip. {In the preceding equations, as well as in the following, we do not write the index $k$ for simplicity of presentation. The operators $\hat{A}(k)$, etc., denote the annihilation operators of the various field modes. The three equations \erefs{eq:BasicRelations}} have a readily apparent physical significance---respectively, they correspond to: the propagation of $\hat{B}_\text{in}$ to reach the particle; the feeding back of $\hat{A}$ to $\hat{C}$ through the ring cavity; and the usual transfer matrix relation for a particle interacting with the four fields surrounding it. The first two of these relations are substituted into the third, which subsequently simplifies to
\begin{equation}
\label{eq:BaseRelation}
\begin{pmatrix}A\\B_\text{in}\end{pmatrix}=\Biggl(\begin{bmatrix}1&0\\0&r_2t_3e^{ik_0l_2}\end{bmatrix}-\hat{M}\begin{bmatrix}\alpha&0\\0&0\end{bmatrix}\Biggr)^{-1}\hat{M}\begin{pmatrix}0\\D\end{pmatrix}
\end{equation}
If we assume far off-resonant operation, \ie, $\partial\zeta/\partial k=0$, the velocity-dependent transfer matrix $\hat{M}$ can be written as~\cite{Xuereb2009b,Xuereb2010b}
\begin{equation}
\begin{bmatrix}1+i\zeta&i\zeta-2i\zeta\tfrac{v}{c}+2ik_0\zeta\tfrac{v}{c}\partial_k\\-i\zeta-2i\zeta\tfrac{v}{c}+2ik_0\zeta\tfrac{v}{c}\partial_k&1-i\zeta\end{bmatrix}\,.
\end{equation}
{The notation $\partial_k\equiv\tfrac{\partial}{\partial k}$ is
used throughout; note that this partial derivative acts not only
on $\alpha(k)$ but also on the field mode amplitudes it precedes.}
\eref{eq:BaseRelation} can be inverted in closed form to first
order in $v/c$, similarly to Ref.~\cite{Xuereb2010b}, and can thus
be used to find $\mathcal{A}=\sqrt{2\epsilon_0S/(\hbar k_0)}\int
A(k)\rmd k$, $\mathcal{B}=\sqrt{2\epsilon_0S/(\hbar k_0)}\int
B(k)\rmd k$, $\mathcal{C}=\sqrt{2\epsilon_0S/(\hbar k_0)}\int
C(k)\rmd k$, and $\mathcal{D}=\sqrt{2\epsilon_0S/(\hbar k_0)}\int
D(k)\rmd k$, {where the normalisation is with respect to} the
pump beam mode area $S$ and {where a monochromatic pump is
assumed}: $B_\text{in}(k)=B_0\delta(k-k_0)${. Here,
$\lvert\mathcal{A}\rvert^2$, $\lvert\mathcal{B}\rvert^2$, etc.,
are the photon currents in units of photons per second}. The
expectation value of the force acting on the particle is then
given by~\cite{Xuereb2009b}:
\begin{equation}
\label{eq:RawForce}
\force_\text{full}=\hbar k_0\bigl(\lvert\mathcal{A}\rvert^2+\lvert\mathcal{B}\rvert^2-\lvert\mathcal{C}\rvert^2-\lvert\mathcal{D}\rvert^2\bigr)\,.
\end{equation}
The values of $\mathcal{A}$, $\mathcal{B}$, etc., from the
solution of \eref{eq:BaseRelation} are then substituted into
\eref{eq:RawForce}, which {we evaluate to} first order in
$v/c$, in terms of $B_0$. After some algebra, we obtain the first
main result of this paper---the friction force acting on the
particle:
\begin{equation}
\label{eq:TMMFriction}
\force=-8\hbar k_0^2\tfrac{v}{c}\re{\frac{\bigl(1-\alpha^\ast\bigr)\zeta\re{\zeta}+i\alpha^\ast\zeta\lvert\zeta\rvert^2}{1-\alpha-i\zeta}\frac{\partial\alpha}{\partial k}}\frac{\bigl|r_2t_3B_0\bigr|^2}{\bigl|1-\alpha-i\zeta\bigr|^2}\,.
\end{equation}
By extending the TMM appropriately, one can keep track of the various noise modes interacting with the system. {\erefs{eq:BasicRelations} then become}
\begin{subequations}
\begin{equation}
\hat{A}=\frac{i\zeta}{1-i\zeta}\hat{B}+\frac{1}{1-i\zeta}\hat{C}+\epsilon\hat{E}\,,
\end{equation}
\begin{equation}
\hat{B}=r_2t_3e^{ik_0l_2}\hat{B}_\text{in}+t_2t_3e^{ik_0l_\text{A}}\hat{E}_\text{isolator}+r_3e^{ik_0l_3}\hat{E}_3^\prime\,,\
\end{equation}
\begin{equation}
\hat{C}=\alpha\hat{A}+r_1e^{ik_0(L-l_1)}\hat{E}_1+t_1r_2\gamma e^{ik_0(L-l_2)}\hat{E}_2+t_1t_2r_3\gamma e^{ik_0(L-l_3)}\hat{E}_3+t_1\epsilon_\text{A}e^{ik_0(L-l_\text{A})}\hat{E}_\text{A}\,, \text{ and}
\end{equation}
\begin{equation}
\hat{D}=\frac{1}{1-i\zeta}\hat{B}+\frac{i\zeta}{1-i\zeta}\hat{C}+\epsilon\hat{E}\,,
\end{equation}
\end{subequations}
with $\epsilon=\sqrt{1-\bigl(1+\lvert\zeta\rvert\bigr)/\lvert
1-i\zeta\rvert^2}$~\cite{Xuereb2009b} and
$\epsilon_\text{A}=\sqrt{1-1/\lvert\gamma\rvert^2}$~\cite{Gardiner2004}.
These equations can be solved simultaneously for $\hat{A}$,
$\hat{B}$, $\hat{C}$, and $\hat{D}$, and the solution used to
evaluate the momentum diffusion constant, $\diffn$, {defined
as} the two-time autocorrelation function of the force operator,
to obtain
\begin{align}
\label{eq:TMMDiffusion}
\diffn\,\delta(t-t^\prime)=2\epsilon_0S\,\hbar k_0\Bigl(&\bigl[\hat{A}(t),\hat{A}^\dagger(t^\prime)\bigr]\lvert\mathcal{A}\rvert^2+\bigl[\hat{B}(t),\hat{B}^\dagger(t^\prime)\bigr]\lvert\mathcal{B}\rvert^2\nonumber\\
&+\bigl[\hat{C}(t),\hat{C}^\dagger(t^\prime)\bigr]\lvert\mathcal{C}\rvert^2+\bigl[\hat{D}(t),\hat{D}^\dagger(t^\prime)\bigr]\lvert\mathcal{D}\rvert^2\nonumber\\
&+2\,\mathrm{Re}\Bigl\{\bigl[\hat{A}(t),\hat{B}^\dagger(t^\prime)\bigr]\mathcal{A}^\ast\mathcal{B}-\bigl[\hat{A}(t),\hat{C}^\dagger(t^\prime)\bigr]\mathcal{A}^\ast\mathcal{C}-\bigl[\hat{A}(t),\hat{D}^\dagger(t^\prime)\bigr]\mathcal{A}^\ast\mathcal{D}\nonumber\\
&\qquad\qquad-\bigl[\hat{B}(t),\hat{C}^\dagger(t^\prime)\bigr]\mathcal{B}^\ast\mathcal{C}-\bigl[\hat{B}(t),\hat{D}^\dagger(t^\prime)\bigr]\mathcal{B}^\ast\mathcal{D}\nonumber\\
&\qquad\qquad+\bigl[\hat{C}(t),\hat{D}^\dagger(t^\prime)\bigr]\mathcal{C}^\ast\mathcal{D}\Bigr\}\Bigr)\,,
\end{align}
keeping in mind that most of the noise modes, as well as $\hat{B}_\text{in}$, obey the commutation relation $\bigl[\hat{E}(t),\hat{E}^\dagger(t^\prime)\bigr]=\hbar k_0/(2\epsilon_0 S)\,\delta(t-t^\prime)$. The sole exception is the noise introduced by the amplifier, $\hat{E}_\text{A}$, for which $\bigl[\hat{E}_\text{A}(t),\hat{E}_\text{A}^\dagger(t^\prime)\bigr]=-\hbar k_0/(2\epsilon_0 S)\,\delta(t-t^\prime)$; this is due to the model of the amplifier as a negative temperature heat-bath, whereby the creation and annihilation operators effectively switch r\^oles. Further discussion of this model can be found in Ref.~\cite[\textsection 7.2]{Gardiner2004}. All the noise modes are independent from one another and from $\hat{B}_\text{in}$, which simplifies the expressions considerably.
\par
Finally, the fluctuation--dissipation theorem~\cite{Metcalf2003} can be used in conjunction with \erefs{eq:TMMFriction} and~(\ref{eq:TMMDiffusion}) to estimate the equilibrium temperature that the motion of the {particle} will tend to:
\begin{equation}
\label{eq:GeneralTemperature}
k_\text{B}T_\text{A}=\frac{\diffn}{-\force/v}\,,
\end{equation}
where $k_\text{B}$ is the Boltzmann constant.

\subsection{The good-cavity limit as a simplified case}\label{sec:GoodCavity}
{Before discussing the result of \erefs{eq:TMMFriction}
and~(\ref{eq:TMMDiffusion}), cf. \Sref{sec:Results}, we shall}
make several approximations to obtain a transparent set of
equations to briefly explore the equilibrium behaviour of the
particle {and to compare with a standard master equation
approach}. In particular, $\zeta$ is assumed to be real, which is
tantamount to assuming that the particle suffers no optical
absorption, i.e., if it is an atom, that it is pumped far
off-resonance. Moreover, the cavity is assumed to be very good
($\lvert t_{1,2,3}\rvert\to 1$) {and thus no gain medium is
introduced in the cavity ($\gamma=1$)}. With these
simplifications, \eref{eq:TMMFriction} reduces to
\begin{align}
\label{eq:TMMFrictionSimple}
\force&\approx -8\hbar k_0^2\tfrac{v}{c}\frac{\zeta^2}{\lvert 1-\alpha\rvert^4}\re{\bigl(1-\alpha^\ast\bigr)^2\frac{\partial\alpha}{\partial k}}\bigl|r_2B_0\bigr|^2\nonumber\\
&\approx 16\hbar k_0^2\zeta^2 v\frac{\kappa\Delta_\text{C}}{\bigl(\Delta_\text{C}^2+\kappa^2\bigr)^2}\frac{1}{\tau}\bigl|r_2B_0\bigr|^2\,.
\end{align}
In the preceding equations, $\Delta_\text{C}$ is the detuning {of the pump} from cavity resonance, $\kappa$ is the HWHM cavity linewidth,
\begin{equation}
\kappa=\frac{1}{\tau}\frac{1-\lvert t_1t_2t_3\rvert\gamma}{\sqrt{\lvert t_1t_2t_3\rvert\gamma}}\,,
\end{equation}
{and} $\tau = L/c$ is the round-trip time{.} Using the same
approximations {as for \eref{eq:TMMFrictionSimple}}, we also
obtain the diffusion constant
\begin{equation}
\label{eq:TMMDiffusionSimple}
\diffn\approx 8\hbar^2 k_0^2\zeta^2\frac{\kappa}{\Delta_\text{C}^2+\kappa^2}\frac{1}{\tau}\bigl|r_2B_0\bigr|^2\,.
\end{equation}
{Note that} $\gamma=1$ {here and therefore}
$\hat{E}_\text{A}$ does not contribute to the diffusion
constant{.} {\erefs{eq:TMMFrictionSimple}
and~(\ref{eq:TMMDiffusionSimple}) hold for the case where
$\Delta_\text{C}/\kappa$ is not too large. The cavity can be fully
described by means of $\kappa$ and the finesse $\mathcal{F}=\pi
c/(2L\kappa)$. Let us now set $\Delta_\text{C}=-\kappa$ in
\erefs{eq:TMMFrictionSimple} and~(\ref{eq:TMMDiffusionSimple}),
whereby
\begin{equation}
\force=-\tfrac{8}{\pi}\hbar k_0^2\zeta^2 v\frac{\mathcal{F}}{\kappa}\bigl|r_2B_0\bigr|^2\,,\text{ and }
\diffn=\tfrac{8}{\pi}\hbar^2 k_0^2\zeta^2\mathcal{F}\bigl|r_2B_0\bigr|^2\,.
\end{equation}
These two expressions have a readily-apparent physical
significance; at a constant finesse, decreasing the cavity
linewidth by making the cavity longer is equivalent to increasing
the retardation effects that underlie this cooling
mechanism~\cite{Xuereb2009a}, leading to a stronger friction
force. At the same time, this has no effect on the intracavity
field strength and therefore does not affect the diffusion. On the
other hand, improving the cavity finesse by reducing losses at the
couplers increases the intracavity intensity, thereby increasing
both the friction force and the momentum diffusion.}
\par
Using \eref{eq:GeneralTemperature} together with
\erefs{eq:TMMFrictionSimple} and~(\ref{eq:TMMDiffusionSimple}) we
obtain, for $\Delta_\text{C}<0$,
\begin{equation}
\label{eq:TMMTemperatureSimple}
T_\text{A}\approx \frac{\hbar}{k_\text{B}}\biggl(\frac{\lvert\Delta_\text{C}\rvert}{\kappa}+\frac{\kappa}{\lvert\Delta_\text{C}\rvert}\biggr)\frac{\kappa}{2}\geq\frac{\hbar}{k_\text{B}}\kappa\,,
\end{equation}
with the minimum temperature occurring at $\Delta_\text{C}=-\kappa$. One notes that this expression is identical to the corresponding one for standard cavity--mediated cooling~\cite{Horak1997}, and can be interpreted in a similar light as the Doppler temperature, albeit with the energy dissipation process shifted from the decay of the atomic excited state to the decay of the cavity field{.}
\par
A particular feature to note in all the preceding expressions is
that they are not spatial averages over the position of the
particle, but they do not depend on this position either. As a
result of this, the force, momentum diffusion and equilibrium
temperature do not in any way depend on the position of the
particle along the cavity field in a 1D model. The issue of
sub-wavelength modulation of the friction force is a major
limitation of cooling methods based on intracavity standing
fields, in particular mirror--mediated cooling~\cite{Xuereb2010b}
and cavity--mediated cooling~\cite{Xuereb2010c}.

\subsection{Comparison with a semiclassical model}\label{sec:Semiclassical}
In the good-cavity limit {and without gain our TMM model is equivalent to a standard master equation approach} with the Hamiltonian
\begin{align}
\label{eq:Hamiltonian}
\hat{H}=&-\hbar\Delta_\text{a}\hat{\sigma}^+\hat{\sigma}^-
-\hbar\Delta_\text{C}\hat{a}_\text{C}^\dagger \hat{a}_\text{C}
\nonumber\\ &+\hbar g\bigl(\hat{a}_\text{C}^\dagger\hat{\sigma}^-
e^{ik_0x}+\hat{\sigma}^+\hat{a}_\text{C}e^{-ik_0x}\bigr)+\hbar
g\bigl(a_\text{P}^\ast\hat{\sigma}^-
e^{-ik_0x}+\hat{\sigma}^+a_\text{P}e^{ik_0x}\bigr)\,,
\end{align}
and the Liouvillian terms
\begin{equation}
\label{eq:Liouvillian}
\mathcal{L}\hat{\rho}=-\Gamma\bigl(\hat{\sigma}^+\hat{\sigma}^-\hat{\rho}-2\hat{\sigma}^-\hat{\rho}\hat{\sigma}^++\hat{\rho}\hat{\sigma}^+\hat{\sigma}^-\bigr)-\kappa\bigl(\hat{a}_\text{C}^\dagger \hat{a}_\text{C}\hat{\rho}-2\hat{a}_\text{C}\hat{\rho} \hat{a}_\text{C}^\dagger+\hat{\rho} \hat{a}_\text{C}^\dagger \hat{a}_\text{C}\bigr)\,,
\end{equation}
as adapted from Ref.~\cite{Gangl2000a} and modified for a
unidirectional cavity where only the unpumped mode is allowed to
circulate{. Here,} $\hat{\rho}$ is the density matrix of the
system, $g$ the atom--field coupling strength, $\hat{a}_\text{C}$
the annihilation operator of the cavity field, $\hat{\sigma}^+$
the atomic dipole raising operator, $\Delta_\text{a}$ the detuning
from atomic resonance, $\Gamma$ the atomic upper state HWHM
linewidth, and $x$ the coordinate of the atom inside the cavity.
The pump field is assumed to be unperturbed by its interaction
with the atom, and in the above is replaced by a c-number,
$a_\text{P}$. {Calculating the friction force form this model
leads again to \eref{eq:TMMFrictionSimple}, thus confirming our
TMM results by a more standard technique. The advantage of the TMM
approach lies in the simplicity and generality of expressions such
as \eref{eq:RawForce}, and the ease with which more optical
elements can be introduced into the system. As shown above,
\eref{eq:TMMDiffusion}, the momentum diffusion coefficient is
easily calculated from the TMM.}

\begin{figure}[t]
 \centering
 \includegraphics[angle=-90,width=0.45\textwidth]{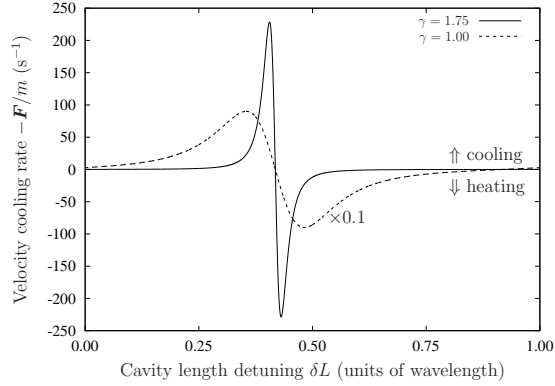}
  \caption{Cooling rate $-(\mathrm{d}v/\mathrm{d}t)/v$ for $^{85}$Rb pumped $-10\Gamma$ from D$_2$ resonance inside a ring-cavity with a round-trip length $L=300$\,m, for two different values of the amplifier gain $\gamma$. Note that the curve for {$\gamma=1$}, as drawn, is scaled \emph{up} by a factor of $10$. The cavity waist is taken to be $10$\,$\upmu$m. ($\lvert t_1\rvert^2=\lvert t_3\rvert^2=0.5$, $\lvert t_2\rvert^2=0.99$, {$B_0$ is chosen such as to give an} atomic saturation $s=0.1$.)}
 \label{fig:Friction}
\end{figure}
\begin{figure}[t]
 \centering
 \includegraphics[angle=-90,width=0.45\textwidth]{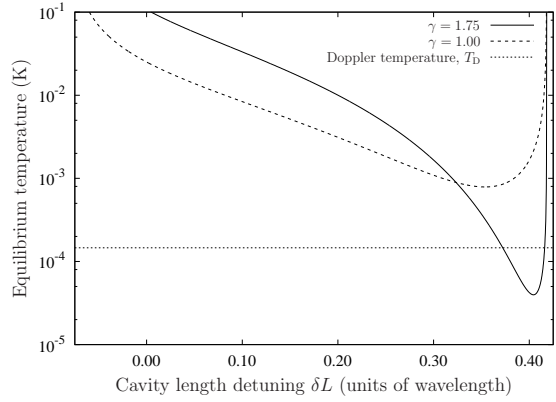}
  \caption{Equilibrium temperature predicted by the transfer matrix model for two values of the amplifier gain $\gamma$. The Doppler temperature for $^{85}$Rb is also indicated. The horizontal axis differs from that in \fref{fig:Friction} mainly because the temperature is only well-defined for regions where the friction force promotes cooling. (Parameters are as in \fref{fig:Friction}.)}
 \label{fig:Temperature}
\end{figure}

\section{Numerical results and discussion}\label{sec:Results}

We can use the conversion factor $\lvert B_0\rvert^2=P/(\hbar
k_0c)$, where $P$ is the power of the input beam, to evaluate the
above equations [notably \erefs{eq:TMMFriction}
and~(\ref{eq:TMMDiffusion})] numerically in a physically
meaningful way. Specifically, the particle is now assumed to be a
(two--level) \textsuperscript{85}Rb atom, pumped $-10\Gamma$ from
D$_2$ resonance, {where} $\Gamma=2\pi\times 3.03$\,MHz {is}
the HWHM linewidth of this same transition at a wavelength of
ca.~$780$\,nm; {because} the detuning is much larger than the
linewidth, {we simplify the calculations} by setting
$\partial\zeta/\partial\omega=0$. The beam waist where the
particle interacts with the field is taken to be $10$\,$\upmu$m.
With the parameters in \fref{fig:Friction}, the power is reduced
by a factor of $1/\lvert t_1t_2t_3\rvert^2=4.04$ with each
round-trip, in the presence of no gain in the amplifier. We shall
compare this case to the low-gain case; the gain of the amplifier
we consider is constrained to be small enough that
$\lvert\alpha\rvert^2=\lvert t_1t_2t_3\gamma|^2<1$. Under these
conditions, there is no {exponential} build-up of intensity
inside the cavity and the system is stable. A cavity with a large
enough gain that $\lvert\alpha\rvert^2>1$ would effectively be a
laser cavity. Such a system {would have} no stable state in our
model, since we assume that the gain medium is not depleted, and
will therefore not be considered further in this paper.
\par
\fref{fig:Friction} shows the friction force acting on the
particle, and \fref{fig:Temperature} the equilibrium temperature,
as the length of the cavity is tuned on the scale of one
wavelength. In each of these two figures two cases are shown, one
representing no gain in the amplifier ({$\gamma=1$}) and one
representing a low-gain amplifier ($\gamma=1.75$); note that in
both cases the condition $\lvert\alpha\rvert^2<1$ is satisfied.

In order to provide a fair comparison between these two cases, we
choose the pump amplitude $B_\text{in}$ such that the saturation
of the particle is the same in the two cases. This ensures that
any difference in cooling performance is not due to a simple
increase in intensity. Since the TMM as presented here is based on
a \emph{linear} model of the particle, our results presented above
are only valid in the limit of saturation parameter much smaller than $1$.
Thus, as a basis for the numerical comparisons between the two
different cases, we choose to set the saturation parameter to
$0.1$. \fref{fig:Friction} shows that under these conditions the
amplified system leads to a significant, approximately $25$-fold,
enhancement of the maximum friction force. This can therefore be
attributed unambiguously to the effective enhancement of the
cavity $Q$-factor by the amplifier.

However, for the parameters considered here, in particular for
small particle polarisability $\zeta$ and for
$\lvert\alpha\rvert^2<1$, the counterpropagating mode intensity is
much smaller than that of the pumped mode, even if the former is
amplified. Thus, the intracavity field is always dominated by the
pump beam, whereas the friction force is mostly dependent on the
Doppler-shifted reflection of the pump from the particle.
Specifically, for the parameters used above we find that the total
field intensity changes by less than 1\% when the gain is
increased from 1 to 1.75. Hence, similar results to those of
\fref{fig:Friction} are obtained even \emph{without} pump
normalisation.

The steady-state temperature, obtained by the ratio of diffusion and friction, \eref{eq:GeneralTemperature}, is shown in \fref{fig:Temperature} for the same parameters as above. We observe that the broader resonance in the friction as a function of cavity detuning (i.e., of cavity length), shown in \fref{fig:Friction}, also leads to a wider range of lower temperatures compared to the amplified case. However, as expected, within the narrower resonance of the amplified system where the friction is significantly enhanced, the stationary temperature is also significantly reduced. We see that while the maximum friction force is increased by a factor of $25.4$, the lowest achievable temperature is decreased by a factor of $19.9$ when switching from $\gamma=1$ to $\gamma=1.75$. While the overall cavity intensity is dominated by the pump field, and is therefore hardly affected by the amplifier, the diffusion is actually dominated by the interaction of the weak counterpropagating field with the pump field. This can be seen most clearly by the strong detuning dependence of the analytic expression for $\diffn$ in the good-cavity limit, \eref{eq:TMMDiffusionSimple}. As a consequence, the lowest achievable temperature is improved by a slightly smaller factor than the maximum friction coefficient. This is consistent with the idea that the amplifier not only increases the cavity lifetime, but also adds a small amount of additional noise into the system. Nevertheless, a strong enhancement of the cooling efficiency is observed in the presence of the amplifier.

\section{Conclusions and Outlook}\label{sec:Conclusions}
We have presented a modified model for optomechanics inside ring cavities where \emph{only one} of the counter-propagating fields in the cavity is allowed to circulate. By pumping the \emph{other} mode and using a gain medium inside the cavity, one can greatly improve the optomechanical force acting on a polarisable particle inside the cavity, regardless of its energy level structure, without bringing about ill effects such as saturation or mirror burning. The conceptual introduction of a gain medium inside the cavity brings about several interesting possibilities. We have considered using this gain medium to offset losses inherent in the cavity, thereby improving its $Q$-factor significantly. This renders possible the use of optical fibres to build the cavity. One could also envisage using doped fibre amplifiers~\cite{Becker1999} to provide a distributed gain medium along the cavity. In this paper, we only considered low-gain media, such that the total losses in the cavity still exceeded the gain. Higher gains could be used to explore and exploit novel phenomena such as optomechanical interactions of weakly reflective micromirrors \emph{inside laser cavities} and will be the subject of future work.

\section*{Acknowledgements}
This work was supported by the UK EPSRC (EP/E039839/1 and EP/E058949/1), and by the Cavity--Mediated Molecular Cooling collaboration within the EuroQUAM programme of the ESF.

\bibliographystyle{tMOP}

\end{document}